\documentclass[letterpaper, 10 pt, conference]{ieeeconf}

\IEEEoverridecommandlockouts

\usepackage{graphics}
\usepackage{graphicx}
\usepackage{epsfig}
\usepackage{amssymb}
\usepackage{amsmath}

\usepackage{amsthm}
\usepackage{mathrsfs}

\usepackage{enumerate}
\usepackage{booktabs}
\usepackage{longtable}
\usepackage{stackengine}
\usepackage{xfrac}
\usepackage{tabularx}
\usepackage{multirow}
\usepackage{cite}
\usepackage[caption=false]{subfig}
\usepackage{float} 
\usepackage{cancel}
\usepackage[normalem]{ulem} 
\usepackage{bm}
\usepackage{soul} 

\usepackage[linesnumbered,ruled,vlined,algo2e,norelsize]{algorithm2e}
\SetArgSty{textnormal}

\usepackage[integrals]{wasysym}

\usepackage{threeparttable}
\usepackage{optidef}
\def\matt#1{\begin{bmatrix}#1\end{bmatrix}}

\makeatletter
\let\NAT@parse\undefined
\makeatother
\usepackage[hidelinks]{hyperref}
\hypersetup{
    colorlinks=true,
    linkcolor=blue,
    filecolor=magenta,
    urlcolor=cyan,
    citecolor=green
}

\usepackage[dvipsnames]{xcolor,colortbl}
\newtheorem{theorem}{Theorem}

\newtheorem{lemma}{Lemma}

\newtheorem{remark}{Remark}

\allowdisplaybreaks

\begin{document}

\allowdisplaybreaks

\title{Real-Time Reference Shaping for Servo Systems}
\author{Zehui Lu, Yebin Wang, Go Sato, and Fujita Tomoya
\thanks{This manuscript is an extended version of a paper accepted for
presentation at the 2026 IEEE Conference on Decision and Control. It includes additional technical details and results that are not included in the conference version due to space limitations. © 2026 IEEE.  Personal use of this material is permitted.  Permission from IEEE must be obtained for all other uses, in any current or future media, including reprinting/republishing this material for advertising or promotional purposes, creating new collective works, for resale or redistribution to servers or lists, or reuse of any copyrighted component of this work in other works.}
\thanks{Z. Lu was with Mitsubishi Electric Research Laboratories (MERL), Cambridge, MA 02139, USA. 
} 
\thanks{Y. Wang is with MERL. {\tt\small Email: yebinwang@ieee.org}}
\thanks{G. Sato and T. Fujita are with the Advanced R\&D Center, Mitsubishi Electric Corporation, Amagasaki, Japan.}
}

\maketitle


\begin{abstract}
This paper addresses real-time motion planning for servo systems subject to nonlinear, state-dependent actuator constraints. A reference reshaping method is proposed that combines analytical optimality with high computational efficiency. Using Karush–Kuhn–Tucker conditions, the problem structure is explicitly characterized, and it is shown that the optimal solution lies within a finite set of candidate points. The complete solution set is constructed via closed-form expressions and a small-scale eigenvalue problem, yielding a deterministic algorithm that recovers the exact optimal solution without iterative optimization or root-finding.
To address kinematic mismatch induced by aggressive commands, a real-time trajectory compensator is introduced to correct accumulated position error while preserving feasibility. Simulation results demonstrate significant computational speed improvements over existing methods, enabling real-time implementation at frequencies approaching 100 kHz.
\end{abstract}

\IEEEpeerreviewmaketitle

\section{Introduction}\label{sec:introduction}

Servo system performs high-speed point-to-point positioning and has been widely used in manufacturing such as Computer Numerical Control (CNC) machines, robotic manipulators, and lithography systems. Its motion control architecture consists of two major stages: trajectory generation \cite{nguyen2008algorithms,wang2013hamiltonian,wang2014real,lam2012model,liu2022constrained} and motor tracking control \cite{reed2016simultaneous,liu2022constrained}. First, a motion planner generates a reference trajectory that drives the work tool to a desired terminal position; and then, a tracking controller regulates motor currents and voltages to track the reference trajectory.

In principle, one could generate dynamically feasible trajectories by solving an optimal control problem that explicitly incorporates actuator dynamics and physical constraints \cite{wang2013hamiltonian,wang2014real, 9050637}. Such approaches are often computationally expensive and difficult to deploy on industrial motion systems that require real-time operation with limited computational resources. As a result, practical systems frequently rely on simplified trajectory generation methods, such as S-curve profiles \cite{nguyen2008algorithms} or bang-bang acceleration strategies \cite{9050637}, that ignore detailed actuator dynamics or constraints. Although computationally efficient, these methods may produce reference trajectories that violate actuator limits, such as torque, current, or voltage constraints \cite{lu24RefReshape}. With such infeasible trajectories being fed into the tracking controller, significant tracking errors or constraint violations may occur. A common remedy is to conservatively limit the allowable acceleration or velocity in the planner, and thus sacrifices productivity.

A recent approach \cite{lu24RefReshape} addresses this challenge by introducing a reference reshaper (RR) that modifies an aggressive reference trajectory to satisfy actuator torque constraints while maintaining computational efficiency. The reshaping problem is formulated as a constrained optimization that projects the desired acceleration onto the dynamically feasible set.
While this framework improves tracking performance and avoids conservative trajectory generation, its computational efficiency remains limited. In particular, the reshaping algorithm requires solving nonlinear equations associated with the constraint boundaries, which introduces nontrivial computational overhead and restricts the implementation to approximately 1~kHz. This limitation makes it challenging to deploy in applications requiring higher planning frequencies.

To overcome this computational bottleneck, a real-time reference reshaping algorithm is developed that achieves both analytical optimality and high computational efficiency. Instead of relying on iterative root-finding to solve nonlinear constraint equations, the proposed method analytically characterizes the structure of the constraint boundaries. It is shown that the optimal solution lies within a finite set of candidate points, which can be constructed through closed-form expressions together with a small-scale eigenvalue computation of a $6 \times 6$ matrix. This analytical characterization leads to two key advantages. First, the complete set of feasible candidates can be explicitly enumerated, enabling systematic selection of the physically admissible solution. Second, the resulting algorithm avoids initialization-dependent iterative procedures and admits deterministic computational complexity. Theoretical guarantees are established through Lemmas~\ref{lemma:root_eq}-\ref{lemma:solve_root} and Theorem~\ref{theorem:reference_reshaper}, which show that the proposed procedure recovers the exact optimal solution of the original problem.
In practice, this formulation yields substantial computational gains, achieving 337× speedup compared to prior reference reshaping methods and 1067× compared to numerical optimization, enabling real-time operation at frequencies approaching 100 kHz.

In addition, a real-time trajectory compensator is introduced to address kinematic mismatch between desired and feasible trajectories. While reference reshaping guarantees dynamic feasibility, aggressive motion commands may introduce accumulated position error due to modified acceleration profiles. The proposed compensator exploits the symmetry of time-optimal trajectories and inserts a constant-velocity phase to recover position deviation online without violating actuator constraints.

The proposed approach can be interpreted as an intermediate layer between motion planning and low-level motor control. Unlike classical reference governor methods \cite{garone2017reference,di2018cascaded,balula2024data}, which enforce constraint satisfaction based on closed-loop system dynamics, the proposed reshaper operates directly on a desired motion trajectory and explicitly incorporates actuator torque–speed limits derived from motor physics.

This leads to a distinct problem structure: the reshaped trajectory must simultaneously preserve kinematic objectives (e.g., reaching the target position $p_{\mathrm{T}}$) and satisfy actuator-level feasibility constraints. The proposed method, therefore, combines elements of trajectory optimization and reference governing, while remaining specifically designed for high-frequency real-time implementation.

The contributions of this paper are summarized as follows:
\begin{itemize}
\item Exact and computationally efficient algorithm: A deterministic reference reshaping algorithm is developed that constructs and evaluates all candidate solutions via closed-form expressions and eigenvalue computation. It is proven that the algorithm recovers the exact optimal solution.
\item Real-time trajectory compensation: A compensation strategy is introduced to correct kinematic mismatch induced by aggressive motion commands while preserving actuator feasibility.
\item High-frequency real-time performance: The proposed method achieves orders-of-magnitude (hundreds to over a thousand times) speed improvements over existing approaches, enabling implementation at frequencies approaching 100 kHz.
\end{itemize}

\begin{figure*}
\centering
\includegraphics[trim=0.1cm 5.9cm 0.6cm 3.7cm, clip, width=0.75\linewidth]{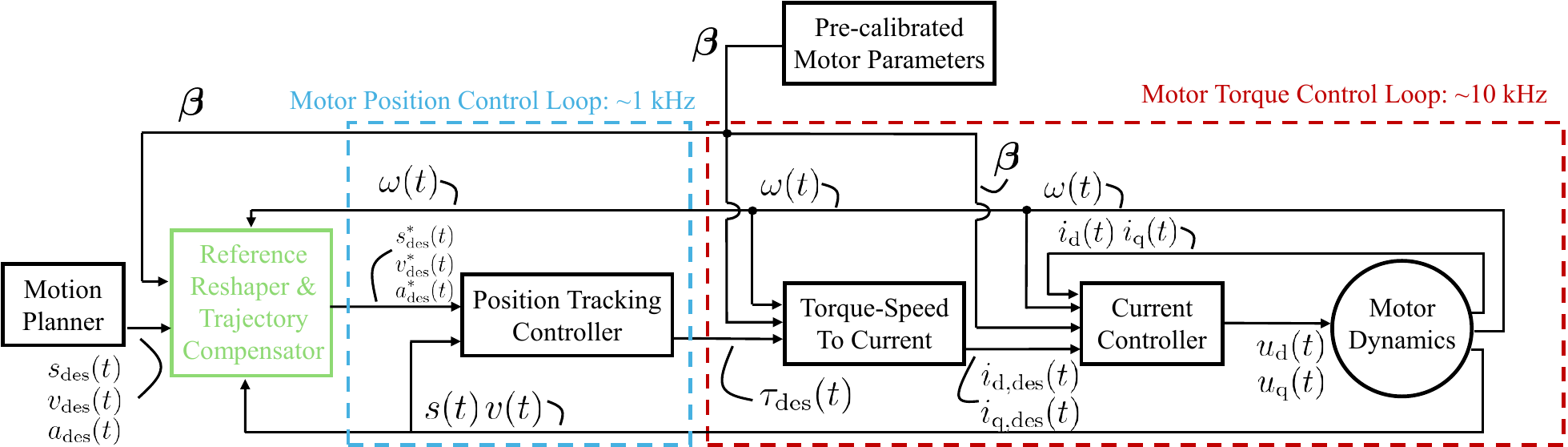}
\caption{The proposed planning and closed-loop control system.} \label{fig:proposed_diagram}
\end{figure*} 

\noindent \textbf{Notations.} For a matrix $\boldsymbol{A} \in \mathbb{R}^{m \times n}$, $\boldsymbol{A}[i,j]$ is the element on the $i$-th row and the $j$-th column, where $i=1, \cdots, m$ and $j=1, \cdots, n$. $\mathbb{Z}_+$ denotes positive integer.

\section{Problem Formulation} \label{sec:problem}

Fig.~\ref{fig:proposed_diagram} illustrates the proposed planning and control architecture. A motion planner generates a desired trajectory for the work tool to reach the target position $p_{\mathrm{T}}$, typically without explicitly accounting for actuator dynamics or physical limits. The proposed RR serves as an intermediate module between the motion planner and the motor tracking controller. It adjusts the desired acceleration based on the current system state to ensure that the resulting trajectory satisfies actuator torque-speed constraints while remaining close to the original reference. The reshaped trajectory is then tracked by the motor controller through the regulation of currents and voltages.

This paper next focuses on the physical system underlying this architecture. Specifically, the work tool is driven by a Surface-Mounted Permanent Magnet Synchronous Motor (SPMSM) and moves along a one-dimensional axis. The reference reshaping problem is formulated based on this actuator-level model, which captures the relationship between motor inputs and the resulting motion. The servo system dynamics are described as follows.
\begin{equation}\label{eq:spmsm_dyn}
\dot{\boldsymbol{x}} = \matt{ \dot i_{\mathrm{d}} \\ \dot i_{\mathrm{q}} \\ \dot{\theta} \\ \dot \omega \\ \dot{s} \\ \dot{v} } = \matt{ \frac{1}{L_{\mathrm{d}}}(-R i_{\mathrm{d}} + p \omega L_{\mathrm{q}} i_{\mathrm{q}} + u_{\mathrm{d}}) \\ \frac{1}{L_{\mathrm{q}}}(-R i_{\mathrm{q}} -(L_{\mathrm{d}} i_{\mathrm{d}} + \Phi_{\mathrm{pm}})p\omega + u_{\mathrm{q}}) \\ \omega \\  1.5 p\Phi_{\mathrm{pm}}i_{\mathrm{q}} J^{-1} \\ v \\  1.5p\Phi_{\mathrm{pm}}i_{\mathrm{q}} J^{-1}\eta },
\end{equation}
where $\boldsymbol{x} \triangleq \matt{ i_{\mathrm{d}} & i_{\mathrm{q}} & \theta & \omega & s & v }^\top \in \mathbb{R}^6$ is the system state, $\boldsymbol{u} \triangleq \matt{ u_{\mathrm{d}} & u_{\mathrm{q}} }^\top \in \mathbb{R}^2$ is the control input and $\eta>0$ is a known gear ratio such that that $v = \eta\omega$. Table~\ref{table:meaning} summarizes the notations used in this paper. The system is subject to the following constraints
\begin{subequations} \label{eq:system_constraint}
\begin{align}
& i_{\mathrm{d}}^2 + i_{\mathrm{q}}^2 \leq I_{\mathrm{max}}^2, \ u_{\mathrm{d}}^2 + u_{\mathrm{q}}^2 \leq V_{\mathrm{dq,max}}^2, \label{eq:system_constraint:current_voltage} \\
& -\omega_{\mathrm{m}}(\boldsymbol{\beta}) \leq \omega \leq \omega_{\mathrm{m}}(\boldsymbol{\beta}), \label{eq:system_constraint:omega}
\end{align}
\end{subequations}
where $V_{\mathrm{dq,max}} \triangleq V_{\mathrm{max}} / \sqrt{3} - R I_{\mathrm{max}} > 0$ is the maximum voltage drop to overcome back-EMF (counter-electromotive force); $\omega_{\mathrm{m}}(\boldsymbol{\beta})$ is the motor's max angular velocity, which is a function of motor parameters $\boldsymbol{\beta}$.
For the remainder of this paper, we denote $\tau \triangleq 1.5p\Phi_{\mathrm{pm}}i_{\mathrm{q}}$ as the motor torque and $a \triangleq\dot{v} = \eta \tau J^{-1} $ as the work tool acceleration.

\begin{table}
\centering
\begin{threeparttable}
\caption{Symbols used in the servo system model} \label{table:meaning}
\begin{tabular}{l l | l l}
\toprule
Symbol & Description & Sym. & Description \\
\midrule
$L_{\mathrm{d}},L_{\mathrm{q}}$ & inductance in d- and q-axis & $\theta, \omega$ & $\dagger$ \\
$i_{\mathrm{d}}, i_{\mathrm{q}}$ & current in d- and q-axis & $s$ & work tool position \\
$u_{\mathrm{d}}, u_{\mathrm{q}}$ & voltage in d- and q-axis & $v$ & work tool velocity \\
$\Phi_{\mathrm{pm}}$ & permanent magnet flux & $J$ & rotor + load inertia \\
$V_{\mathrm{max}}$ & max DC bus voltage & $p$ & number of pole pairs \\
$I_{\mathrm{max}}$ & max current & $R$ & winding resistance \\
\bottomrule
\end{tabular}
\begin{tablenotes}
\small
\item $\dagger$: rotor angular mechanical position and velocity, respectively
\end{tablenotes}
\end{threeparttable}
\centering
\end{table}

Based on this model, the problem of interest is to design a real-time motion planning and control strategy that drives the work tool to the target position $p_{\mathrm{T}} \in \mathbb{R}$ as quickly as possible, while ensuring that the resulting motion remains dynamically feasible with respect to actuator limits.

\section{Preliminaries} \label{sec:exist}

This section presents the modeling of motor torque capacity and reviews the existing reference reshaping approach in \cite{lu24RefReshape}, which serves as the foundation for the proposed method.

\subsection{Motor Torque Capacity Modeling} \label{subsec:torque_capacity}

\begin{figure}
\centering
\includegraphics[width=0.49\linewidth]{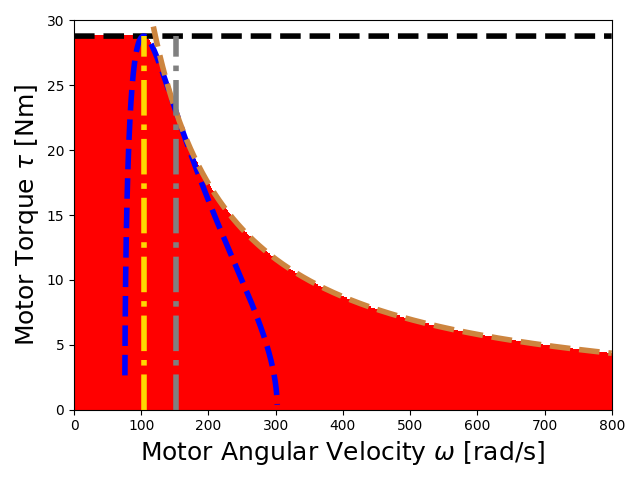}
\caption{Motor torque bound for case $\Phi_{\mathrm{pm}}/L_{\mathrm{d}} < I_{\mathrm{max}}$. The red and white region represents the feasible and infeasible operation range, respectively.} \label{fig:torque_map}
\end{figure}

Given the max torque per ampere control (MTPA) strategy \cite{consoli2009steady} for a motor, together with the sign of $\Phi_{\mathrm{pm}}/L_{\mathrm{d}} - I_{\mathrm{max}}$ \cite{lu2025differentiable}, the torque bound is given by piecewise analytic functions of speed.
The max torque function with $\Phi_{\mathrm{pm}}/L_{\mathrm{d}} - I_{\mathrm{max}} < 0$ is exemplified as \eqref{eq:motor_torque_max_negative}:
\begin{equation} \label{eq:motor_torque_max_negative}
\!\!\!\!\tau_{\mathrm{m}}(\omega, \boldsymbol{\beta}) = \begin{cases}
\frac{3}{2} p \Phi_{\mathrm{pm}}I_{\mathrm{max}},&\text{if }  \vert \omega \vert  \leq \omega_{\mathrm{r}}(\boldsymbol{\beta}) , \\
\frac{3}{2}p \Phi_{\mathrm{pm}}i_{\mathrm{q,lim}},&\text{if }  \vert \omega \vert  \in [\omega_{\mathrm{r}}(\boldsymbol{\beta}),\omega_{\mathrm{s}}(\boldsymbol{\beta})], \\
\frac{3}{2}p \Phi_{\mathrm{pm}}i_{\mathrm{q,lim,V}},&\text{if }  \vert \omega \vert  \in [\omega_{\mathrm{s}}(\boldsymbol{\beta}),\infty) ,
\end{cases}
\end{equation}
where $\omega_{\mathrm{r}}(\boldsymbol{\beta})$ and $\omega_{\mathrm{s}}(\boldsymbol{\beta})$ are given by
\begin{equation*}
\omega_{\mathrm{r}} = 
\textstyle\frac{V_{\mathrm{dq,max}}}{p\sqrt{(L_{\mathrm{q}}I_{\mathrm{max}})^2+\Phi_{\mathrm{pm}}^2}}, \omega_{\mathrm{s}} = \frac{V_{\mathrm{dq,max}}}{p\sqrt{(L_{\mathrm{d}} I_{\mathrm{max}})^2 - \Phi_{\mathrm{pm}}^2}};
\end{equation*}
$i_{\mathrm{q,lim,V}}(\omega, \boldsymbol{\beta})$ and $i_{\mathrm{q,lim}}(\omega, \boldsymbol{\beta})$ are given by 
\begin{align*}
& i_{\mathrm{q,lim,V}} = V_{\mathrm{dq,max}} / (p\omega L_{\mathrm{q}}), \  i_{\mathrm{q,lim}} = (I_{\mathrm{max}}^2 - i_{\mathrm{d,lim}}^2)^{0.5}, \\
&i_{\mathrm{d,lim}} = \textstyle\frac{ (V_{\mathrm{dq,max}}/(p\omega))^2 - (L_{\mathrm{d}} I_{\mathrm{max}})^2 - \Phi_{\mathrm{pm}}^2 }{2\Phi_{\mathrm{pm}}L_{\mathrm{d}}}.
\end{align*}
This piecewise-defined function is visualized in Fig. \ref{fig:torque_map}. The yellow and gray dash-dot vertical lines represent $\omega_{\mathrm{r}}(\boldsymbol{\beta})$ and $\omega_{\mathrm{s}}(\boldsymbol{\beta})$, respectively. The black dashed horizontal line, the blue and brown dashed curves correspond to the first, second, and third rows of \eqref{eq:motor_torque_max_negative}, respectively. Regardless of the sign of $\Phi_{\mathrm{pm}}/L_{\mathrm{d}} - I_{\mathrm{max}}$, there always exists following properties:
\begin{enumerate}
\item $\tau_{\mathrm{m}}(\omega,\boldsymbol{\beta})$ is a constant max torque $\tau_{\mathrm{m,c}}(\boldsymbol{\beta}) \triangleq 1.5p \Phi_{\mathrm{pm}}I_{\mathrm{max}}$ when $\omega \in [0, \omega_{\mathrm{r}}(\boldsymbol{\beta})]$;
\item $\tau_{\mathrm{m}}(\omega,\boldsymbol{\beta})$ is monotonically decreasing when $\omega \in [\omega_{\mathrm{r}}(\boldsymbol{\beta}), \omega_{\mathrm{m}}(\boldsymbol{\beta})]$; $\tau_{\mathrm{m}} = 0$ when $\omega = \omega_{\mathrm{m}}$;
\item $\omega_{\mathrm{m}}(\boldsymbol{\beta}) > \omega_{\mathrm{r}}(\boldsymbol{\beta})$ is the max motor speed and $\omega_{\mathrm{m}}(\boldsymbol{\beta}) = \infty$ when $\Phi_{\mathrm{pm}}/L_{\mathrm{d}} - I_{\mathrm{max}} \leq 0$.
\item when $\omega < 0$, $\tau_{\mathrm{m}}(\omega,\boldsymbol{\beta}) = \tau_{\mathrm{m}}(-\omega,\boldsymbol{\beta})$.
\end{enumerate}
Given motor speed $\omega$, desired torque $\tau_{\mathrm{des}}$ and analytical torque bound such as \eqref{eq:motor_torque_max_negative}, the desired motor currents $i_\mathrm{d,des},i_\mathrm{q,des}$ can be computed analytically.
\cite[Algorithm~1]{lu2025differentiable} provides the detailed steps of this analytical method, which is referred to as ``Torque-Speed To Current'' in Fig.~\ref{fig:proposed_diagram}.

\begin{figure}
\centering
\includegraphics[width=0.49\linewidth]{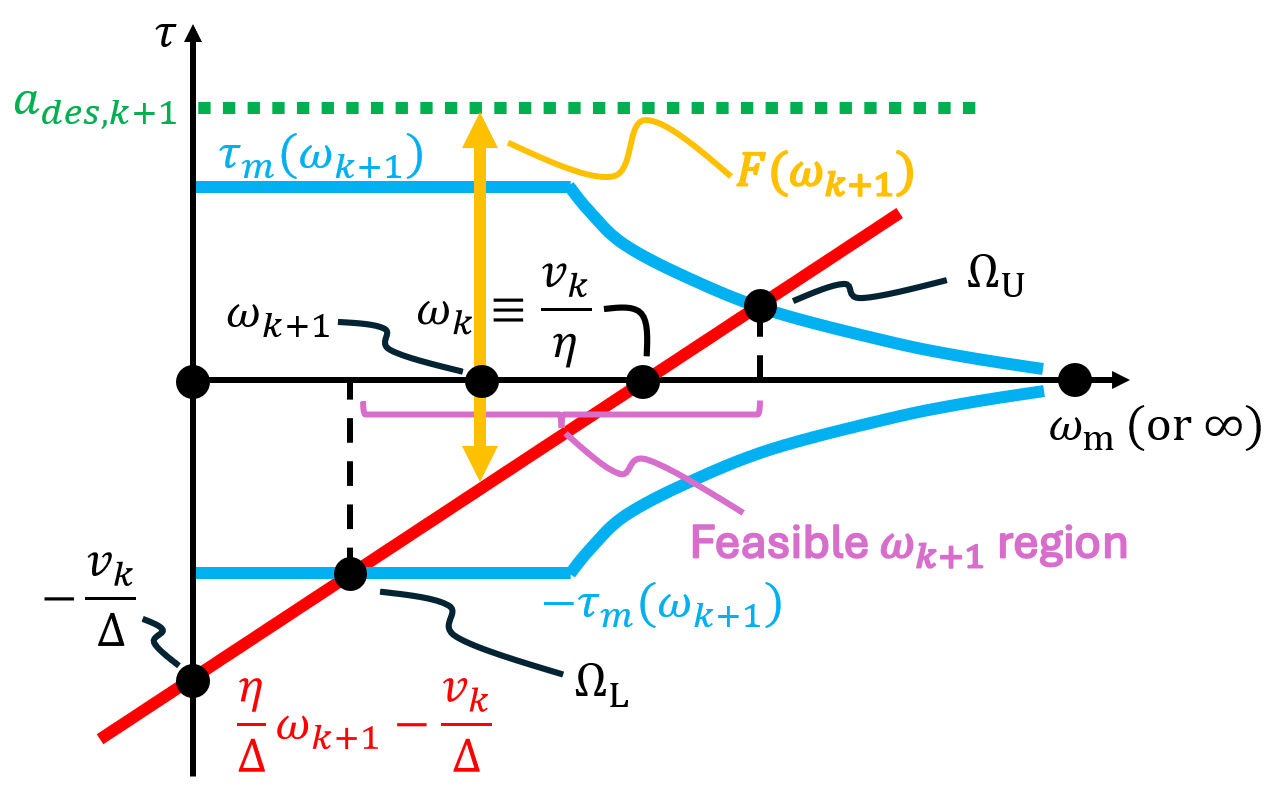}
\caption{A geometric illustration of the optimization \eqref{acc_tracking:reparam:entire}.} \label{fig:acc_tracking_geometry}
\end{figure}

\subsection{Existing Reference Shaping Algorithm} \label{ssec:prelim:rr}

Suppose the RR runs for every time interval $\Delta > 0$. At each timestamp $t_k$, given the desired acceleration (from the motion planner) $a_{\mathrm{des},k+1} \triangleq a_{\mathrm{des}}(t_{k+1})$ for duration $[t_k, t_{k+1} = t_k + \Delta)$, the RR solves the following constrained optimization
\begin{mini!}|s|
{\omega_{k+1}}{ F(\omega_{k+1}):= ||(\frac{\eta}{\Delta}\omega_{k+1} - \frac{v_k}{\Delta}) - a_{\mathrm{des},k+1}||^2 \label{acc_tracking:reparam}}
{\label{acc_tracking:reparam:entire}}{}
\addConstraint{ |\frac{J}{\Delta} \omega_{k+1} - \frac{Jv_k}{\Delta \eta}| \leq \gamma \tau_{\mathrm{m}}(\omega_{k+1})\label{acc_tracking:primal_1}}
\addConstraint{ -\omega_{\mathrm{m}}(\boldsymbol{\beta}) \leq\omega_{k+1}\leq \omega_{\mathrm{m}}(\boldsymbol{\beta}), \label{acc_tracking:primal_2}}
\end{mini!}
where $\gamma \in (0,1]$ is a prescribed torque margin constant near 1, permitting a slight overshoot during the transient of trajectory tracking. If the optimal motor speed from \eqref{acc_tracking:reparam:entire} is given by $\omega_{k+1}^*$, then the optimal desired motions are $a^*_{\mathrm{des},k+1} = (\eta\omega^*_{k+1}-v_k)/\Delta$, $v^*_{\mathrm{des},k+1} = \eta\omega^*_{k+1}$, and $s^*_{\mathrm{des},k+1} = s_k + v_k \Delta + 0.5 a^*_{\mathrm{des},k+1} \Delta^2$.

Unlike the classic reference/command governor that solves a constrained optimization for every reference point with a maximal constraint admissible set defined explicitly \cite{di2018cascaded}, the RR searches over a finite set of critical points for the optimal solution. By checking the Karush–Kuhn–Tucker (KKT) necessary conditions on the optimization problem \eqref{acc_tracking:reparam:entire}, one can search over only a finite set of critical points (at most six) of motor speed for the optimal solution of \eqref{acc_tracking:reparam:entire}.
To begin with, the optimal $w_{k+1}^*$ must either be a stationary point of the objective function \eqref{acc_tracking:reparam}, or lie on the boundary of the feasible region. 
If $w_{k+1}^*$ is a stationary point of the objective function, it must be $\omega_{k+1}^* = (v_k+\Delta a_{\mathrm{des},k+1})/\eta$.
On the other hand, if $w_{k+1}^*$ lies on the boundary of the feasible region, then $w_{k+1}^* \in \Omega_{\mathrm{L}} \cup \Omega_{\mathrm{U}} \cup \{ \pm \omega_{\mathrm{m}}(\beta)\}$, where $\Omega_{\mathrm{L}}$ and $\Omega_{\mathrm{U}}$ are the roots of constraints \eqref{acc_tracking:primal_1} when they are active:
\begin{subequations} \label{eq:root_eq}
\begin{align}
\Omega_{\mathrm{L}} &\triangleq \{\omega \ | \ \xi_{\mathrm{L}}(\omega) := \gamma \tau_{\mathrm{m}}(\omega)+(\frac{J}{\Delta}\omega - \frac{J v_k}{\Delta \eta})=0\}, \\
\Omega_{\mathrm{U}} &\triangleq \{\omega \ | \ \xi_{\mathrm{U}}(\omega) := \gamma \tau_{\mathrm{m}}(\omega)-(\frac{J}{\Delta}\omega - \frac{J v_k}{\Delta \eta}) = 0\}.
\end{align}
\end{subequations}

From a geometric perspective, Fig.~\ref{fig:acc_tracking_geometry} illustrates the optimization \eqref{acc_tracking:reparam:entire}. The slope and the intercept of the red line are fixed given a particular $v_k$. The possible $a_{k+1}$ can be any points along the red line segment intersected with the torque bound $\tau_{\mathrm{m}}(\omega_{k+1})$ and $-\tau_{\mathrm{m}}(\omega_{k+1})$. The optimal $a_{k+1}^*$ is the point with the minimal distance to the desired acceleration $a_{\mathrm{des},k+1}$ (in green dotted line).
Since the slope $\eta/\Delta$ of the red line cannot be zero, $\Omega_{\mathrm{L}}$ and $\Omega_{\mathrm{U}}$ must contain only one point, respectively.

\section{Real-Time Reference Reshaping and Trajectory Compensation} \label{sec:reference_reshaper}

This section first introduces the improved reference reshaping algorithm that resolves the computational bottleneck of the prior approach. Then, a trajectory compensator is presented, which augments the reshaped trajectory to correct position deviations in real time.

\subsection{Improved Algorithm for Real-time Reference Reshaping}

As discussed in Sec.~\ref{ssec:prelim:rr}, the primary computational bottleneck of the RR lies in evaluating the constraint boundaries $\Omega_{\mathrm{U}}$ and $\Omega_{\mathrm{L}}$. From \eqref{eq:root_eq}, this task reduces to finding all real roots of a nonlinear equation of $\omega$:
\begin{equation}
\gamma \tau_{\mathrm{m}}(\omega,\boldsymbol{\beta}) \pm (\frac{J}{\Delta}\omega - \frac{J v_k}{\Delta \eta})=0.
\end{equation}
Regardless of the range of $\omega$ and the sign of $\Phi_{\mathrm{pm}}/L_{\mathrm{d}} - I_{\mathrm{max}}$, $\tau_{\mathrm{m}}(\omega, \boldsymbol{\beta})$ has three function branches at most, as shown in \eqref{eq:motor_torque_max_negative}.
Thus, Lemma~\ref{lemma:root_eq} presents all the possible upper and lower bound roots, $\omega_{\mathrm{U}}$ and $\omega_{\mathrm{L}}$.
A proof is provided in Appendix~\ref{appendix:proof_lemma1}.

\begin{lemma} \label{lemma:root_eq}
The equation $\gamma \tau_{\mathrm{m}}(\omega) \pm (\frac{J}{\Delta}\omega - \frac{J v_k}{\Delta \eta})=0$ at most has three branches for its roots $\omega$, regardless of whether the roots are real and in the branch-associated range.

\noindent 1) For the first branch of \eqref{eq:motor_torque_max_negative},
\begin{subequations}
\begin{align}
\omega_{\mathrm{U}} &= \omega_k + 1.5p \Phi_{\mathrm{pm}}I_{\mathrm{max}} \Delta \gamma / J, \\
\omega_{\mathrm{L}} &= \omega_k - 1.5p \Phi_{\mathrm{pm}}I_{\mathrm{max}} \Delta \gamma / J.
\end{align}
\end{subequations}
2) For the third branch,
\begin{subequations} \label{eq:thrid_branch}
\begin{align}
&\omega_{\mathrm{U}} = 0.5\omega_k \pm 0.5 \sqrt{\bar{{\Delta}}_\mathrm{U}}, \ \omega_{\mathrm{L}} = 0.5\omega_k \pm 0.5 \sqrt{\bar{{\Delta}}_\mathrm{L}}, \label{eq:thrid_branch:1} \\
&\bar{{\Delta}}_\mathrm{U} \triangleq \omega_k^2 + 6 \gamma \Phi_{\mathrm{pm}} V_{\mathrm{dq,max}} \Delta / (L_{\mathrm{q}} J), \label{eq:thrid_branch:2} \\
&\bar{{\Delta}}_\mathrm{L} \triangleq \omega_k^2 - 6 \gamma \Phi_{\mathrm{pm}} V_{\mathrm{dq,max}} \Delta / (L_{\mathrm{q}} J). \label{eq:thrid_branch:3}
\end{align}
\end{subequations}
3) For the second branch, for both bounds, the roots $\omega$ satisfy 
\begin{subequations} \label{eq:case2_poly_3_new}
\begin{align}
&\omega^6 + a_5(\boldsymbol{\beta}) \omega^5 + a_4(\boldsymbol{\beta}) \omega^4 + a_2(\boldsymbol{\beta}) \omega^2 + a_0(\boldsymbol{\beta}) = 0, \label{eq:case2_poly_3_new:main}\\
&a_0(\boldsymbol{\beta}) \triangleq \frac{ 1.5^2 \gamma^2 \Delta^2  V_{\mathrm{dq,max}}^4 }{ 4J^2p^2L_{\mathrm{d}}^2 }, \quad a_5(\boldsymbol{\beta}) \triangleq -2 \omega_k, \\
&a_2(\boldsymbol{\beta}) \triangleq -\frac{ 1.5^2 V_{\mathrm{dq,max}}^2 (L_{\mathrm{d}}^2 I_{\mathrm{max}}^2 + \Phi_{\mathrm{pm}}^2)\gamma^2 \Delta^2 }{ 2 L_{\mathrm{d}}^2 J^2 }, \\
&a_4(\boldsymbol{\beta}) \triangleq \omega_k^2 + \frac{ (L_{\mathrm{d}}^2 I_{\mathrm{max}}^2 - \Phi_{\mathrm{pm}}^2)^2 1.5^2 p^2 \gamma^2\Delta^2 }{ 4 L_{\mathrm{d}}^2 J^2 }.
\end{align}
\end{subequations}

\end{lemma}

Note that \eqref{eq:case2_poly_3_new} is a 6th-degree polynomial, which has no analytical formula for roots by the Abel–Ruffini theorem. And Lemma~\ref{lemma:root_eq} does not include the feasibility check on the real domain and the range of $\Omega_{\mathrm{U}}$ and $\Omega_{\mathrm{L}}$ in each branch. Therefore, Lemma~\ref{lemma:solve_root} provides a computationally efficient way to calculate $\Omega_{\mathrm{U}}$ and $\Omega_{\mathrm{L}}$. A proof is provided in Appendix~\ref{appendix:proof_lemma2}.

\begin{lemma} \label{lemma:solve_root}
Define a companion matrix of \eqref{eq:case2_poly_3_new} as:
\begin{equation}\label{eq:companion_main}
\!\!\!\!\!\! \boldsymbol{C}(\boldsymbol{\beta}) \triangleq \matt{ 0 & 1 & 0 & 0 & 0 & 0 \\ 0 & 0 & 1 & 0 & 0 & 0 \\ 0 & 0 & 0 & 1 & 0 & 0 \\ 0 & 0 & 0 & 0 & 1 & 0 \\ 0 & 0 & 0 & 0 & 0 & 1 \\ -a_0(\boldsymbol{\beta}) & 0 & -a_2(\boldsymbol{\beta}) & 0 & -a_4(\boldsymbol{\beta}) & -a_5 }.
\end{equation}
Applying Algorithm~\ref{alg:lower_upper} yields the appropriate $\Omega_{\mathrm{L}}$ and $\Omega_{\mathrm{U}}$ defined in \eqref{eq:root_eq}.
\end{lemma}

\begin{algorithm2e}
\DontPrintSemicolon
\KwIn{$\omega_k, \boldsymbol{\beta}, p, \Delta, \gamma, J, V_{\mathrm{dq,max}},I_{\mathrm{max}}, \Omega_{\mathrm{L}} = \Omega_{\mathrm{U}} = \emptyset$}

$\omega_{\mathrm{r}} \gets \textstyle\frac{V_{\mathrm{dq,max}}}{p\sqrt{(\hat{L}_{\mathrm{q}}I_{\mathrm{max}})^2+\Phi_{\mathrm{pm}}^2}}$, \ 
$\omega_{\mathrm{L}} \gets \omega_k - \frac{1.5p \Phi_{\mathrm{pm}}I_{\mathrm{max}} \Delta \gamma }{ J } $, \ $\omega_{\mathrm{U}} \gets \omega_k + \frac{1.5p \Phi_{\mathrm{pm}}I_{\mathrm{max}} \Delta \gamma }{ J } $\;

\lIf{$|\omega_{\mathrm{L}}| \leq \omega_{\mathrm{r}}$}
{$\Omega_{\mathrm{L}} \gets \Omega_{\mathrm{L}} \cup \{ \omega_{\mathrm{L}} \}$}
\lIf{$|\omega_{\mathrm{U}}| \leq \omega_{\mathrm{r}}$}
{$\Omega_{\mathrm{U}} \gets \Omega_{\mathrm{U}} \cup \{ \omega_{\mathrm{U}} \}$}

$\boldsymbol{C}(\boldsymbol{\beta}) \gets $ \eqref{eq:companion_main} given $\boldsymbol{\beta}$\;
$\Lambda_{\mathrm{L}}, \Lambda_{\mathrm{U}} \gets$ Get the set of real eigenvalues of $\boldsymbol{C}$\;

\uIf{$\Phi_{\mathrm{pm}}/L_{\mathrm{d}} - I_{\mathrm{max}} > 0$}
{
$\omega_{\mathrm{max}} \gets \frac{V_{\mathrm{dq,max}}}{p  \vert \Phi_{\mathrm{pm}}-L_{\mathrm{d}}I_{\mathrm{max}} \vert }$ \;
$\bar{\Lambda}_{\mathrm{L}} \gets \{ \omega \in \Lambda_{\mathrm{L}} \ | \ |\omega| \in [\omega_{\mathrm{r}}, \omega_{\mathrm{max}}]  \}$\;
$\bar{\Lambda}_{\mathrm{U}} \gets \{ \omega \in \Lambda_{\mathrm{U}} \ | \ |\omega| \in [\omega_{\mathrm{r}}, \omega_{\mathrm{max}}]  \}$\;
}
\uElseIf{$\Phi_{\mathrm{pm}}/L_{\mathrm{d}} - I_{\mathrm{max}} = 0$}
{
$\bar{\Lambda}_{\mathrm{L}} \gets \{ \omega \in \Lambda_{\mathrm{L}} \ | \ |\omega| \in [\omega_{\mathrm{r}}, \infty)  \}$\;
$\bar{\Lambda}_{\mathrm{U}} \gets \{ \omega \in \Lambda_{\mathrm{U}} \ | \ |\omega| \in [\omega_{\mathrm{r}}, \infty)  \}$\;
}
\Else{
$\omega_{\mathrm{s}} \gets \frac{V_{\mathrm{dq,max}}}{p\sqrt{(L_{\mathrm{d}} I_{\mathrm{max}})^2 - \Phi_{\mathrm{pm}}^2}}$ \;
$\bar{\Lambda}_{\mathrm{L}} \gets \{ \omega \in \Lambda_{\mathrm{L}} \ | \ |\omega| \in [\omega_{\mathrm{r}}, \omega_{\mathrm{s}}]  \}$\;
$\bar{\Lambda}_{\mathrm{U}} \gets \{ \omega \in \Lambda_{\mathrm{U}} \ | \ |\omega| \in [\omega_{\mathrm{r}}, \omega_{\mathrm{s}}]  \}$\;
}

$\Omega_{\mathrm{L}} \gets \Omega_{\mathrm{L}} \cup \bar{\Lambda}_{\mathrm{L}}$, $\Omega_{\mathrm{U}} \gets \Omega_{\mathrm{U}} \cup \bar{\Lambda}_{\mathrm{U}}$\;

\If{$\Phi_{\mathrm{pm}}/L_{\mathrm{d}} - I_{\mathrm{max}} < 0$}
{
$\bar{\Delta}_{\mathrm{L}} =  \omega_k^2 - 4 \frac{ 1.5 \gamma \Phi_{\mathrm{pm}} V_{\mathrm{dq,max}} \Delta }{ \hat{L}_{\mathrm{q}} J } $\;
\If{$\bar{\Delta}_{\mathrm{L}} \geq 0$}
{
$\Lambda_1 \gets \{ 0.5\omega_k \pm 0.5 \sqrt{\bar{{\Delta}}_{\mathrm{L}}} \}$\;
$\bar{\Lambda}_1 \gets \{ \omega \in \Lambda_1 \ | \ |\omega| \in [\omega_{\mathrm{s}}, \infty)  \}$\;
$\Omega_{\mathrm{L}} \gets \Omega_{\mathrm{L}} \cup \bar{\Lambda}_1$\;
}

$\bar{\Delta}_{\mathrm{U}} =  \omega_k^2 + 4 \frac{ 1.5 \gamma \Phi_{\mathrm{pm}} V_{\mathrm{dq,max}} \Delta }{ \hat{L}_{\mathrm{q}} J } $\;
\If{$\bar{\Delta}_{\mathrm{U}} \geq 0$}
{
$\Lambda_2 \gets \{ 0.5\omega_k \pm 0.5 \sqrt{\bar{{\Delta}}_{\mathrm{U}}} \}$\;
$\bar{\Lambda}_2 \gets \{ \omega \in \Lambda_2 \ | \ |\omega| \in [\omega_{\mathrm{s}}, \infty)  \}$\;
$\Omega_{\mathrm{U}} \gets \Omega_{\mathrm{U}} \cup \bar{\Lambda}_2$\;
}
}
\Return{$\Omega_{\mathrm{L}}$, $\Omega_{\mathrm{U}}$}
\caption{Lower \& Upper Bound Algorithm}
\label{alg:lower_upper}
\end{algorithm2e}

By incorporating Algorithm~\ref{alg:lower_upper}, the computation bottleneck of the original RR is eliminated, leading to the real-time implementation summarized in Algorithm~\ref{alg:rr_improved_main}.
The following theorem guarantees the algorithm's optimality, and the proof is provided in Appendix~\ref{appendix:theorem}.
\begin{theorem} \label{theorem:reference_reshaper}
By calculating the roots searching over the finite set of solution candidates $\Bar \Omega$, the solution returned by Algorithm~\ref{alg:rr_improved_main} is the optimal solution of \eqref{acc_tracking:reparam:entire}.
\end{theorem}

\begin{remark}
A natural approach to solving the nonlinear equation in \eqref{eq:root_eq} is to use iterative root-finding methods such as Newton-Raphson. However, such methods can only converge to a single root depending on initialization and do not provide information about the full set of real roots. This limitation is critical in the reference reshaping problem, where multiple feasible roots may exist. Without access to all candidate solutions, it is not possible to determine which root satisfies the physical constraints.

In contrast, the proposed approach leverages Lemma~\ref{lemma:root_eq} and Lemma~\ref{lemma:solve_root} to explicitly characterize all real roots. Combined with the current motor speed, this enables systematic selection of the physically admissible solution, as implemented in Lines 6-16 of Algorithm~\ref{theorem:reference_reshaper}.
\end{remark}

While the improved RR guarantees dynamic feasibility, it does not explicitly account for the position deviation introduced by modifying the desired acceleration profile. This limitation motivates the development of a trajectory compensation mechanism, presented next.

\begin{algorithm2e}
\DontPrintSemicolon
\KwIn{$\omega_k, v_k, {s_k}, a_{\mathrm{des},k+1}, \boldsymbol{\beta}, p, \Delta, \gamma,\eta,J,V_{\mathrm{dq,max}},I_{\mathrm{max}}$}
$\Omega_{\mathrm{L}}, \Omega_{\mathrm{U}} \gets$ Algorithm~\ref{alg:lower_upper}\;
\uIf{$\Phi_{\mathrm{pm}}/L_{\mathrm{d}} - I_{\mathrm{max}} > 0$}
{
$\omega_{\mathrm{max}} \gets \frac{V_{\mathrm{dq,max}}}{p  \vert \Phi_{\mathrm{pm}}-L_{\mathrm{d}}I_{\mathrm{max}} \vert }$ \;
$\Omega\gets \{ \omega_k + \frac{\Delta a_{\mathrm{des},k+1}}{\eta} \} \cup \Omega_{\mathrm{U}} \cup \Omega_{\mathrm{L}} \cup \{ \pm \omega_{\mathrm{max}}, \omega_k \}$\;
$\bar{\Omega} \gets \{\omega \in \Omega \ | \  |\omega|\leq \omega_{\mathrm{max}}, \  |\frac{J}{\Delta} \omega - \frac{J v_k}{\Delta \eta}| \leq \gamma \tau_{\mathrm{m}}(\omega)\}$\;
}
\Else{
$\Omega\gets \{ \omega_k + \frac{\Delta a_{\mathrm{des},k+1}}{\eta} \} \cup \Omega_{\mathrm{U}} \cup \Omega_{\mathrm{L}} \cup \{ \omega_k \}$\;
$\bar{\Omega} \gets \{\omega \in \Omega \ | \  |\frac{J}{\Delta} \omega - \frac{J v_k}{\Delta \eta}| \leq \gamma \tau_{\mathrm{m}}(\omega)\}$\;
}
$\omega_{k+1}^* {\gets} \underset{\omega\in \Bar{\Omega}}{\operatorname{\arg\min}} \ F(\omega), \ a^*_{\mathrm{des},k+1} \gets (\eta\omega^*_{k+1}-v_k)/\Delta$\;
${v^*_{\mathrm{des},k+1} \gets \eta \omega^*_{k+1},\  s^*_{\mathrm{des},k+1} \gets s_k + \textstyle\frac{(v^*_{\mathrm{des},k+1}+v_k)\Delta}{2}}$\;
\Return{$\omega_{k+1}^*$, $a^*_{\mathrm{des},k+1}$, $v^*_{\mathrm{des},k+1}$, $s^*_{\mathrm{des},k+1}$}
\caption{Real-time Reference Reshaper}
\label{alg:rr_improved_main}
\end{algorithm2e}

\subsection{Trajectory Compensation and Integrated Algorithm}

To address the position deviation introduced by reference reshaping, a real-time trajectory compensator is proposed to adjust the reshaped trajectory online. The design leverages the symmetric structure of time-optimal trajectories, where the acceleration follows a bang-bang profile with a single switching point. This structure provides a natural opportunity to introduce compensation without violating feasibility.

The overall procedure is summarized in Algorithm~\ref{alg:compen}. Let the desired trajectory be 
$\mathcal{T} = \{ t_k, s_{\mathrm{des},k}, v_{\mathrm{des},k}, a_{\mathrm{des},k} \ | \ k = 0,\cdots, T \}$, 
where $s_{\mathrm{des},T} = p_{\mathrm{T}}$, and $T\in \mathbb{Z}_+$ is the terminal index. At each planning cycle, the algorithm updates the reshaped reference for $[t_k, t_{k+1})$ based on the current state at time $t_k$.
To account for the insertion of a compensation phase, the algorithm maintains a time offset $t_{\mathrm{comp}}$ and maps the current time $t_{k+1}$ to the original trajectory time 
$t_{\mathrm{orig},k+1}$ (Line 4). The desired acceleration $a_{\mathrm{des},k+1}$ is then obtained by interpolating $\mathcal{T}$ at $t_{\mathrm{orig},k+1}$ (Line 7), and the RR (Algorithm~\ref{alg:rr_improved_main}) computes the feasible command $a^*_{\mathrm{des},k+1}, v^*_{\mathrm{des},k+1}, s^*_{\mathrm{des},k+1}$ (Line 8).

The trajectory compensator is activated based on the condition in Line 9. Specifically, it detects the switching point of the desired trajectory, indicated by a sign change in $a_{\mathrm{des},k+1}$, together with a mismatch between the reshaped and original accelerations ($a^*_{\mathrm{des},k+1} \neq a_{\mathrm{des},k+1}$). When this condition is satisfied, and no compensation has been applied ($t_{\mathrm{comp},h} = 0$), the algorithm enters the compensation phase by setting the flag to true and recording the current velocity $v_{\mathrm{des},h}$ (Line 10).
During the compensation phase (Lines 11-14), the acceleration command is set to zero, and a constant-velocity motion is enforced with $v_{\mathrm{des},h}$. The position is updated accordingly, while the accumulated compensation time $t_{\mathrm{comp}}$ increases.

The duration of the compensation phase is determined using the trajectory's symmetry. When the position reaches the midpoint $s^*_{\mathrm{des},k+1} = 0.5 p_{\mathrm{T}}$, the algorithm records the half-duration $t_{\mathrm{comp},h}$ (Line 15). The compensation phase terminates when the total compensation time reaches $2\,t_{\mathrm{comp},h}$ (Line 5), after which the trajectory resumes the nominal deceleration phase.
Throughout this process, the mapping between $t_{k+1}$ and $t_{\mathrm{orig},k+1}$ ensures that the RR continues to operate consistently with respect to the original trajectory, despite the inserted constant-velocity segment. As a result, the trajectory compensator corrects the accumulated position deviation while preserving actuator feasibility.

Overall, the integrated algorithm combines reference reshaping and trajectory compensation into a unified real-time procedure. It achieves accurate terminal position tracking without introducing additional trajectory optimization problems, while maintaining computational efficiency suitable for high-frequency implementation.

\begin{algorithm2e}
\DontPrintSemicolon
\KwIn{$\Delta, \gamma,\boldsymbol{\beta}, p,\eta,J,V_{\mathrm{dq,max}},I_{\mathrm{max}}, p_{\mathrm{T}}, t_{\mathrm{T}}, \mathcal{T}$}

$\texttt{flag} = \texttt{False}, \ t_{\mathrm{comp}} = t_{\mathrm{comp,h}} = 0$\;
\While{}{
current time $t_k$, next time $t_{k+1} \gets t_k + \Delta$\;
$t_{\mathrm{orig},k+1} \gets t_{k+1} - t_{\mathrm{comp}}$\;

\lIf{ $t_{\mathrm{comp,h}} > 0$ \&\& $t_{\mathrm{comp}} == 2 t_{\mathrm{comp,h}}$}{$\texttt{flag} \gets \texttt{False}$}

Measure $s_k, v_k, \omega_k$\;
$ a_{\mathrm{des},k+1} \gets$ Interpolate $\mathcal{T}$ with time $t_{\mathrm{orig},k+1}$\;
$a^*_{\mathrm{des},k+1}, v^*_{\mathrm{des},k+1}, s^*_{\mathrm{des},k+1} \gets$ Algorithm~\ref{alg:rr_improved_main}\;
$\texttt{cond} \gets!  \texttt{flag}$ \&\& $(a^*_{\mathrm{des},k+1} != a_{\mathrm{des},k+1})$ \&\& $(a_{\mathrm{des},k+1} \cdot a_{\mathrm{des},k} < 0)$ \&\& $(t_{\mathrm{comp,h}} == 0)$ \;
\lIf{$\texttt{cond}$}{
$\texttt{flag} \gets \texttt{True}, v_{\mathrm{des,h}} \gets v_{\mathrm{des},k}$
}
\If{$\texttt{flag}$}{
$a_{\mathrm{des}, k+1}^* \gets 0, \ v_{\mathrm{des}, k+1}^* \gets v_{\mathrm{des,h}}$\;
$s_{\mathrm{des}, k+1}^* \gets s_{\mathrm{des}, k} + \Delta \cdot v_{\mathrm{des,h}}$\;
$ t_{\mathrm{comp}} \gets t_{\mathrm{comp}} + \Delta$\;
}

\lIf{ $s_{\mathrm{des}, k+1}^* == 0.5 p_{\mathrm{T}}$ }{
$t_{\mathrm{comp, h}} \gets t_{\mathrm{comp}}$
}
Perform motor control until timestamp $t_{k+1}$\;
}
\caption{Real-time Reference Reshaping Process with Trajectory Compensator}
\label{alg:compen}
\end{algorithm2e}

\section{Simulation Results} \label{sec:simulation}

This section presents simulation results to validate the proposed algorithm in terms of dynamic feasibility and computational efficiency.
The system parameters are $p=4$, $\Delta = 1$ ms, $\gamma = 0.97$, $R=0.08\ \Omega$, $J=0.15$ $\text{kg}\cdot\text{m}^2$, $\eta=0.0075$, $I_{\mathrm{max}}=40$ A, and $V_{\mathrm{max}}=100\sqrt{3}$ V. The system dynamics \eqref{eq:spmsm_dyn} are propagated using Euler integration with time step $\Delta_{\mathrm{s}}=0.1$ ms. The true motor torque bound is shown in Fig.~\ref{fig:torque_map}.

\subsection{Position \& Current Tracking Controller}

Let $a_{\mathrm{des}}(t)$, $v_{\mathrm{des}}(t)$, and $s_{\mathrm{des}}(t)$ denote the original desired trajectory, and let $(\cdot)^*$ denote the reshaped trajectory. As shown in Fig.~\ref{fig:proposed_diagram}, the position tracking controller operates at 1 kHz and computes the desired motor torque based on the tracking error:
\begin{subequations}
\begin{align}
\begin{split}
&\tau_{\mathrm{des}}(t) = \frac{J a_{\mathrm{des}}(t)}{\eta} - K_{\mathrm{P},\tau} e_{s}(t) - K_{\mathrm{I},\tau} \int_0^t e_{s}(\tau) d\tau \\
&\quad\quad\quad\quad - K_{\mathrm{D},\tau} e_{v}(t),
\end{split}\\
&e_{s}(t) \triangleq s(t) - s_{\mathrm{des}}(t), \  e_{v}(t) \triangleq v(t) - v_{\mathrm{des}}(t).
\end{align}
\end{subequations}

The current controller operates at 10 kHz. Given the desired torque $\tau_{\mathrm{des}}(t)$, the ``Torque-Speed To Current'' module computes $i_{\mathrm{d,des}}(t)$ and $i_{\mathrm{q,des}}(t)$, and the controller regulates the current tracking errors $e_{i_{\mathrm{d}}}(t)$ and $e_{i_{\mathrm{q}}}(t)$ through voltage inputs:
\begin{subequations}
\begin{align}
&e_{i_{\mathrm{d}}}(t) \triangleq i_{\mathrm{d}}(t) - i_{\mathrm{d,des}}(t), \ e_{i_{\mathrm{q}}}(t) \triangleq i_{\mathrm{q}}(t) - i_{\mathrm{q,des}}(t), \\
\begin{split}
&u_{\mathrm{d,des}}(t) = R i_{\mathrm{d,des}}(t) - L_{\mathrm{q}} p\omega(t) i_{\mathrm{q}}(t) \\
&\quad\  - K_{\mathrm{P},i_{\mathrm{d}}} e_{i_{\mathrm{d}}}(t) - K_{\mathrm{I},i_{\mathrm{d}}} \textstyle\int_0^t e_{i_{\mathrm{d}}}(\tau) d\tau,
\end{split}\\
\begin{split}
&u_{\mathrm{q,des}}(t) = R i_{\mathrm{q,des}}(t) + (L_{\mathrm{d}} i_{\mathrm{d}}(t) + \Phi_{\mathrm{pm}})p\omega(t) \\
&\quad\  - K_{\mathrm{P},i_{\mathrm{q}}} e_{i_{\mathrm{q}}}(t) - K_{\mathrm{I},i_{\mathrm{q}}} \textstyle\int_0^t e_{i_{\mathrm{q}}}(\tau) d\tau.
\end{split}
\end{align}
\end{subequations}

\subsection{Reference Reshaper}

The proposed RR is evaluated in terms of (i) dynamic feasibility and (ii) computational efficiency.

\textbf{Simulation setup:}
The desired motion has a stroke of 3 m. The motion planner generates a time-optimal bang-bang trajectory, where maximum acceleration $a_{\mathrm{m}} = 1.44$ m/s$^2$ and maximum velocity $v_{\mathrm{m}} = 5$ m/s are applied symmetrically.
The following methods are compared: B (baseline without RR), A (original RR in \cite{lu24RefReshape}), C (solving optimization problem \eqref{acc_tracking:reparam:entire} numerically via IPOPT \cite{wachter2006implementation}), and P (proposed RR).

\begin{figure*}
\subfloat[Baseline approach B: Desired and actual torque trajectory without RR.]
{\label{fig:sim:torque_baseline} \includegraphics[width=0.26\linewidth]{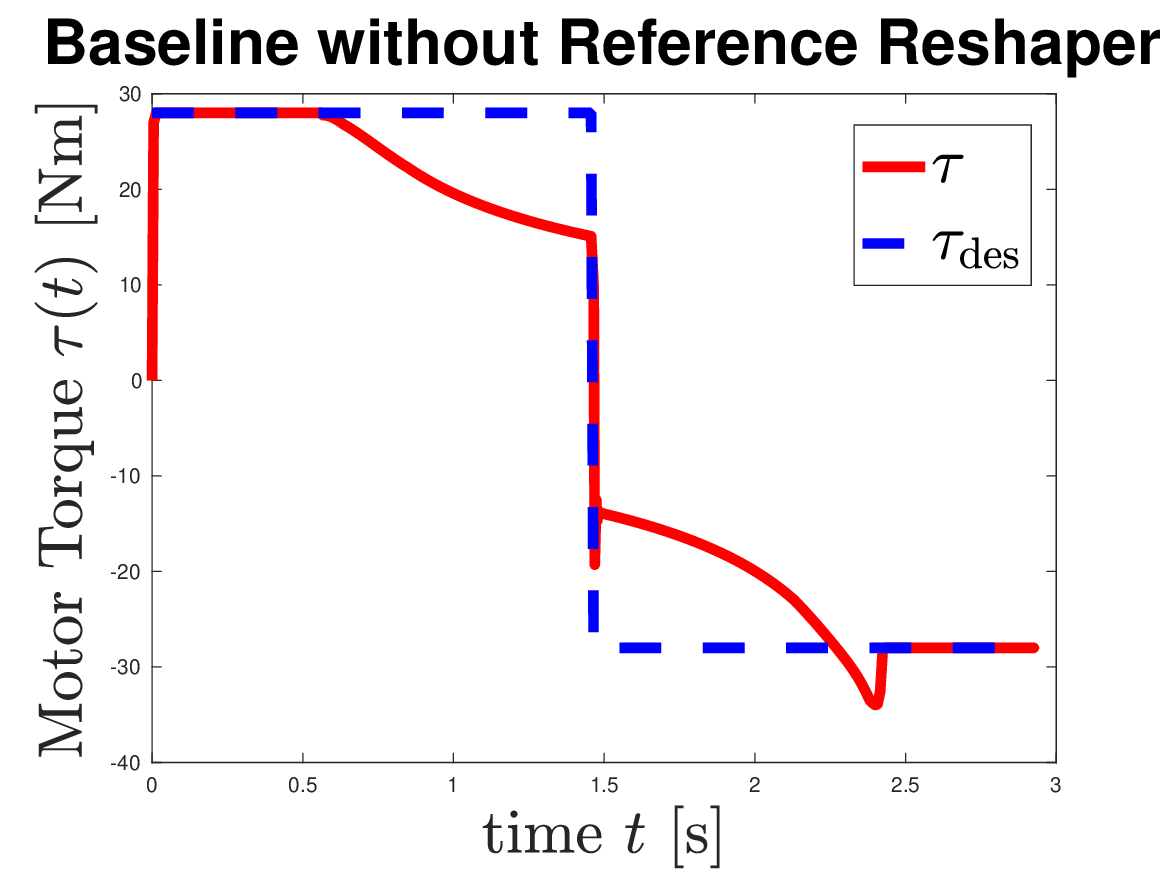}}
\hfill
\subfloat[Desired and actual torque trajectory with RR (P).]
{\label{fig:sim:torque_proposed} \includegraphics[width=0.26\linewidth]{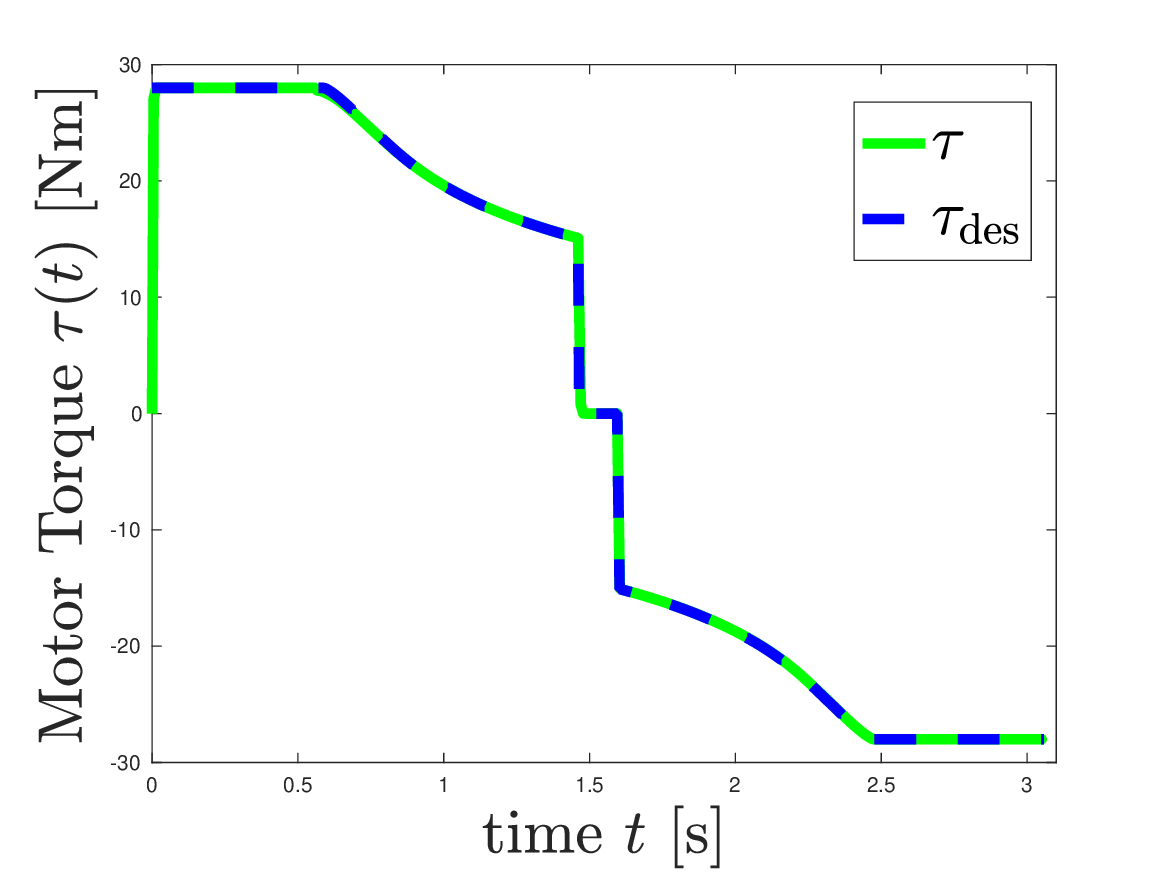}}
\hfill
\subfloat[Desired motion before (with $_{\mathrm{des}}$) and after RR (with $^*$) vs actual motion (P).]
{\label{fig:sim:motion_proposed} \includegraphics[width=0.26\linewidth]{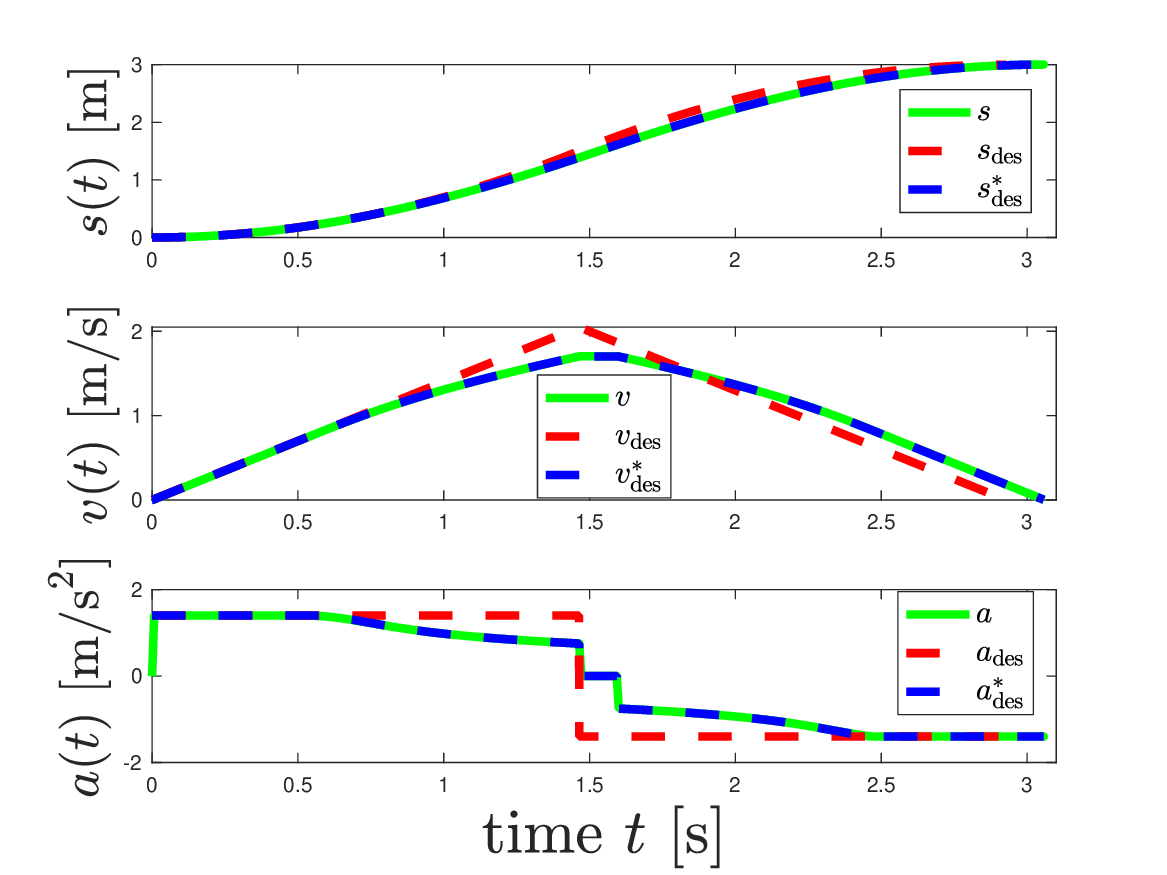}}
\caption{Desired and actual torque and motion trajectory of different methods.} \label{fig:sim}
\end{figure*}

\textbf{Dynamic feasibility:}
Due to the torque-speed characteristics (Fig.~\ref{fig:torque_map}), the aggressive trajectory generated by the motion planner is dynamically infeasible. As shown in Fig.~\ref{fig:sim:torque_baseline}, method B produces torque commands that violate actuator limits, leading to current and voltage constraint violations, as indicated by the torque spike around 2.3 s. Reducing $a_{\mathrm{m}}$ mitigates infeasibility but results in a more conservative trajectory with increased motion time.

In contrast, Fig.~\ref{fig:sim:torque_proposed} shows that the proposed method (P) reshapes the aggressive trajectory into a feasible one. The actual torque closely tracks the reshaped reference, confirming constraint satisfaction. Fig.~\ref{fig:sim:motion_proposed} further demonstrates that the reshaped trajectory remains close to the original desired motion while eliminating infeasibility.
Table~\ref{table:sim} shows that the baseline method results in large position error due to infeasibility, whereas the proposed method achieves accurate tracking (2.4 mm error) without violating constraints and with comparable motion time.

\textbf{Computational efficiency:}
The computation time of different RR implementations is summarized in Table~\ref{table:computation}, on a 2017 MacBook Pro with a 3.1 GHz Intel Core i7 and 16 GB RAM. The proposed method achieves an average computation time of $0.0100 \pm 0.0029$ ms, compared to $3.3743 \pm 2.2564$ ms for method A and $10.6664 \pm 13.0116$ ms for method C.

This corresponds to a $337\times$ speedup over the original RR (A) and a $1067\times$ speedup over the numerical optimization approach (C). The significant reduction in both mean computation time and variance indicates a deterministic and lightweight computational structure.
At this level of efficiency, the reference reshaping algorithm can be executed at frequencies approaching 100 kHz, enabling high-frequency real-time implementation beyond conventional limits (e.g., 1 kHz in prior work).

Overall, the results demonstrate that the proposed method achieves dynamic feasibility and accurate tracking while substantially improving computational efficiency.

\begin{table}
\centering
\begin{threeparttable}
\caption{Performance Summary for Reference Reshaper} \label{table:sim}
\begin{tabular}{c | c c}
\toprule
Performance Index & B & P \\
\midrule
Violated constraints & Y & $\bm{N}$ \\
Motion Time [s] & \sout{2.9277} (Violated constraints) & $\bm{3.0570}$ \\
Final Pos. Error [mm] & 205.1 & $\bm{2.4}$ \\
Avg. RR Comp. Time [ms] & N/A & $\bm{0.0100}$ \\
\bottomrule
\end{tabular}
\end{threeparttable}
\centering
\end{table}

\begin{table}
\centering
\begin{threeparttable}
\caption{Computation Time Statistics for Reference Reshaper} \label{table:computation}
\begin{tabular}{c | c | c }
\toprule
P (Time [ms]) & A (Time [ms]) & C (Time [ms]) \\
\midrule
$\bm{0.0100 \pm 0.0029}$ & 3.3743 $\pm$ 2.2564 & 10.6664 $\pm$ 13.0116 \\
\bottomrule
\end{tabular}
\end{threeparttable}
\centering
\end{table}

\section{Conclusion and Future Work} \label{sec:conclusion}

A reference reshaping framework for motion planning under actuator constraints has been developed, leveraging KKT conditions to replace iterative optimization with exact, deterministic evaluation. This analytical structure yields 337× and 1064× speed improvements over existing methods, enabling dynamically feasible, real-time implementation at frequencies approaching 100 kHz. Future work will investigate robust reference reshaping to account for modeling errors and parameter uncertainties. Specifically, a primary focus will be analyzing how perturbations in critical SPMSM parameters, such as inductance ($L_\mathrm{d}$, $L_\mathrm{q}$) and permanent magnetic flux $\Phi_\mathrm{pm}$, affect the analytical reshaping boundaries, alongside the integration of real-time parameter estimation.


\bibliographystyle{IEEEtran}


\appendix

\subsection{Proof of Lemma~\ref{lemma:root_eq}} \label{appendix:proof_lemma1}

The analysis begins with the root-finding problem associated with the upper bound equation $\gamma \tau_{\mathrm{m}}(\omega,\boldsymbol{\beta}) - (\frac{J}{\Delta}\omega - \frac{J v_k}{\Delta \eta})=0$. The derivation for the lower bound proceeds analogously and is omitted unless distinctions are mentioned.

1) In the first branch, the upper bound root equation is
\begin{equation}
1.5p \gamma \Phi_{\mathrm{pm}}I_{\mathrm{max}}  -\frac{J}{\Delta}\omega + \frac{J v_k}{\Delta \eta} = 0.
\end{equation}
With the fact that $v_k = \eta \omega_k$, $\omega_{\mathrm{U}} = \frac{ v_k + 1.5p \Phi_{\mathrm{pm}}I_{\mathrm{max}} \Delta \gamma \eta/J }{ \eta } = \omega_{\mathrm{U}} = \omega_k + 1.5p \Phi_{\mathrm{pm}}I_{\mathrm{max}} \Delta \gamma/J$.
Similarly, $\omega_{\mathrm{L}} = \omega_k - 1.5p \Phi_{\mathrm{pm}}I_{\mathrm{max}} \Delta \gamma/J$.

2) In the third branch, the upper bound root equation is
\begin{equation}
1.5p \gamma \Phi_{\mathrm{pm}}i_{\mathrm{q,lim,V}} (\omega) - \frac{J}{\Delta}\omega + \frac{J v_k}{\Delta \eta} = 0,
\end{equation}
where $i_{\mathrm{q,lim,V}}(\omega) = V_{\mathrm{dq,max}} / (p\omega L_{\mathrm{q}})$. This equation reads
\begin{subequations} \label{eq:third}
\begin{align}
&1.5p \gamma \Phi_{\mathrm{pm}} \frac{ V_{\mathrm{dq,max}} }{ p L_{\mathrm{q}} } \frac{1}{\omega} - \frac{J}{\Delta}\omega + \frac{J v_k}{\Delta \eta} = 0 \Rightarrow \\
&-\frac{J}{\Delta}\omega^2 + \frac{J v_k}{\Delta \eta} \omega + \frac{ 1.5 \gamma \Phi_{\mathrm{pm}} V_{\mathrm{dq,max}} }{ L_{\mathrm{q}} } = 0 \Rightarrow \\
& \omega^2 - \frac{v_k}{\eta} \omega - \frac{ 1.5 \gamma \Phi_{\mathrm{pm}} V_{\mathrm{dq,max}} \Delta }{ L_{\mathrm{q}} J } = 0. \label{eq:third:final}
\end{align}
\end{subequations}
With $v_k = \eta\omega_k$, check whether \eqref{eq:third:final} has real roots by
\begin{equation} \label{eq:case_2_cond_upper}
\bar{\Delta}_{\mathrm{U}} = \omega_k^2 + 4 \frac{ 1.5 \gamma \Phi_{\mathrm{pm}} V_{\mathrm{dq,max}} \Delta }{ L_{\mathrm{q}} J } \geq 0.
\end{equation}
And the real root is given by
\begin{equation} \label{eq:case_2_upper}
\omega_{\mathrm{U}} = \frac{v_k}{2\eta} \pm \frac{1}{2} \sqrt{\bar{{\Delta}}_{\mathrm{U}}} = \frac{\omega_k}{2} \pm \frac{1}{2} \sqrt{\bar{{\Delta}}_{\mathrm{U}}}.
\end{equation}
Similarly, the upper bound roots yield \eqref{eq:thrid_branch:1} and \eqref{eq:thrid_branch:3}.

In the second branch, the upper bound root equation is
\begin{equation} \label{eq:root_eq_case2_upper}
1.5p \gamma \Phi_{\mathrm{pm}}i_{\mathrm{q,lim}}(\omega) - \frac{J}{\Delta}\omega + \frac{J v_k}{\Delta \eta} = 0,
\end{equation}
Let $a \triangleq \frac{ V_{\mathrm{dq,max}}^2  }{2p^2\Phi_{\mathrm{pm}}L_{\mathrm{d}}}$ and $b \triangleq \frac{ L_{\mathrm{d}}^2 I_{\mathrm{max}}^2 + \Phi_{\mathrm{pm}}^2 }{ 2\Phi_{\mathrm{pm}}L_{\mathrm{d}} }$, \eqref{eq:root_eq_case2_upper} reads
\begin{equation}
1.5p \gamma \Phi_{\mathrm{pm}} \sqrt{ - \frac{a^2}{\omega^4} + \frac{2ab}{\omega^2} + (I_{\mathrm{max}}^2 - b^2) } = \frac{J}{\Delta}\omega -\frac{J v_k}{\Delta \eta},
\end{equation}
which further reads
\begin{subequations} \label{eq:root_equation_2_upper}
\begin{align}
\begin{split}
&\sqrt{ - \frac{a^2}{\omega^4} + \frac{2ab}{\omega^2} + (I_{\mathrm{max}}^2 - b^2) } = \\
&-\frac{J v_k}{1.5p \gamma \Phi_{\mathrm{pm}} \Delta \eta} + \frac{J}{1.5p \gamma \Phi_{\mathrm{pm}}\Delta}\omega
\end{split}\\
&\Rightarrow \sqrt{ - \frac{a^2}{\omega^4} + \frac{2ab}{\omega^2} + (I_{\mathrm{max}}^2 - b^2) } = -c + d\omega = d(\omega - \omega_k), \label{eq:root_equation_2_upper:b}
\end{align}
\end{subequations}
where $c \triangleq \frac{J v_k}{1.5p \gamma \Phi_{\mathrm{pm}} \Delta \eta}$ and $d \triangleq \frac{J}{1.5p \gamma \Phi_{\mathrm{pm}}\Delta}$.
Multiplying $\omega^2$ to both sides of \eqref{eq:root_equation_2_upper:b} yields
\begin{equation} \label{eq:case2_poly_1_upper}
\begin{split}
&\sqrt{ - \Bigl(a^2 - 2ab\omega^2 + (b^2 - I_{\mathrm{max}}^2) \omega^4\Bigl) } = -c \omega^2 + d\omega^3 \\
&=\omega^2(d\omega-c) = d\omega^2(\omega-\frac{c}{d}) = d\omega^2(\omega - \omega_k).
\end{split}
\end{equation}
$d\omega^2(\omega - \omega_k)$ has to be non-negative, which means 1) when $\omega_k \geq 0$, feasible solution lies in $[\omega_k, \infty)$; 2) when $\omega_k \leq 0$, feasible solution lies in $[\omega_k, 0]$.

\begin{remark}
For the lower-bound case, the root equation in the second branch leads to
\begin{equation*}
\sqrt{ - \Bigl(a^2 - 2ab\omega^2 + (b^2 - I_{\mathrm{max}}^2) \omega^4\Bigl) } = -d\omega^2(\omega - \omega_k).
\end{equation*}
Thus, $-d\omega^2(\omega - \omega_k)$ has to be non-negative, which means 1) when $\omega_k \geq 0$, feasible solution lies in $[0, \omega_k]$; 2) when $\omega_k \leq 0$, feasible solution lies in $(-\infty, \omega_k]$.
\end{remark}

To calculate the root for the equation \eqref{eq:case2_poly_1_upper}, taking a square for both sides yields
\begin{equation} \label{eq:case2_poly_2_upper}
- a^2 + 2ab\omega^2 + (I_{\mathrm{max}}^2 - b^2) \omega^4 = (d\omega^3-c \omega^2)^2,
\end{equation}
Dividing $d^2$ for both side of \eqref{eq:case2_poly_2_upper} yields \eqref{eq:case2_poly_3_new:main},
where the coefficients are
\begin{subequations} \label{eq:coeff}
\begin{align}
a_0 &= \frac{a^2}{d^2} = (\frac{ 1.5 \gamma \Delta  V_{\mathrm{dq,max}}^2 }{ 2JpL_{\mathrm{d}} })^2 = \frac{ 1.5^2 \gamma^2 \Delta^2  V_{\mathrm{dq,max}}^4 }{ 4J^2p^2L_{\mathrm{d}}^2 }, \\
a_2 &= -\frac{2ab}{d^2} = -\frac{ 1.5^2 V_{\mathrm{dq,max}}^2 (L_{\mathrm{d}}^2 I_{\mathrm{max}}^2 + \Phi_{\mathrm{pm}}^2)\gamma^2 \Delta^2 }{ 2 L_{\mathrm{d}}^2 J^2 }, \\
\begin{split}
a_4 &= \frac{c^2+b^2-I_{\mathrm{max}}^2}{d^2} \\
&= \omega_k^2 + \frac{ (L_{\mathrm{d}}^2 I_{\mathrm{max}}^2 - \Phi_{\mathrm{pm}}^2)^2 1.5^2 p^2 \gamma^2\Delta^2 }{ 4 L_{\mathrm{d}}^2 J^2 } ,
\end{split}\\
a_5 &= -\frac{2c}{d} = -\frac{ 2v_k }{ \eta } = -2 \omega_k.
\end{align}
\end{subequations}

\begin{remark}
The coefficients of the 6th-order polynomial are the same for the lower and upper bound, due to the square operation applied to \eqref{eq:case2_poly_1_upper}. \qed
\end{remark}

\subsection{Proof of Lemma~\ref{lemma:solve_root}} \label{appendix:proof_lemma2}

The root sets $\Omega_{\mathrm{L}}$ and $\Omega_{\mathrm{U}}$ are defined by the nonlinear boundary equation \eqref{eq:root_eq}. The motor torque function $\tau_{\mathrm{m}}(\omega)$ is piecewise analytic in $\omega$ and consists of at most three branches, as characterized in \eqref{eq:motor_torque_max_negative}, Sec.~\ref{subsec:torque_capacity}. The proof proceeds by analyzing each branch separately.

For the first and third branches, Lemma~\ref{lemma:root_eq} provides explicit closed-form expressions for all candidate roots. The admissible roots are obtained by enforcing the corresponding range constraints on $\omega$. These steps are implemented in Lines 1-3 and 18-28 of Algorithm~\ref{alg:lower_upper}, which construct the corresponding subsets of $\Omega_{\mathrm{L}}$ and $\Omega_{\mathrm{U}}$.

For the second branch, the boundary condition reduces to the sixth-order polynomial \eqref{eq:case2_poly_3_new}. All real roots of this polynomial can be obtained by computing the real eigenvalues of the associated companion matrix in \eqref{eq:companion_main}. The valid roots are then selected by applying the same range constraints on $\omega$. This procedure is summarized in Lines 4-16 of Algorithm~\ref{alg:lower_upper}.
Combining the results from all three branches yields the complete sets $\Omega_{\mathrm{L}}$ and $\Omega_{\mathrm{U}}$. \qed

\subsection{Proof of Theorem~\ref{theorem:reference_reshaper}} \label{appendix:theorem}

By Lemma~\ref{lemma:root_eq} and Lemma~\ref{lemma:solve_root}, Algorithm~\ref{alg:lower_upper} computes the sets $\Omega_{\mathrm{L}}$ and $\Omega_{\mathrm{U}}$, which characterize all boundary points of $\omega$ satisfying the constraint \eqref{eq:root_eq}. These boundary points are provided for Algorithm~\ref{alg:rr_improved_main} (Line 1).
Algorithm~\ref{alg:rr_improved_main} then augments these boundary points with additional candidate solutions to form a finite candidate set $\bar{\Omega}$, as constructed in Lines 5 and 8. 
From \cite[Theorem~1]{lu24RefReshape}, the optimal solution of the optimization problem \eqref{acc_tracking:reparam:entire} is guaranteed to lie within this finite set $\bar{\Omega}$. Therefore, evaluating the objective over $\bar{\Omega}$ and selecting the minimizer (Lines 9-11 of Algorithm~\ref{alg:rr_improved_main}) yields the optimal solution.
\qed

\end{document}